%% file: Dengue-tweets-classification-SoWeMine.tex
\documentclass{llncs}

\input{preamble}

\begin{document}

\title{Tracking Dengue Epidemics using Twitter Content Classification and Topic Modelling}
\author{Paolo Missier\inst{1} 
  		\and Alexander Romanovsky\inst{1} 
  		\and Tudor Miu\inst{1} 
  		\and Atinder Pal\inst{1}
  		\and Michael Daniilakis\inst{1}
  		\and Alessandro Garcia\inst{2}
  		\and Diego Cedrim\inst{2}
  		\and Leonardo da Silva Sousa\inst{2}
  		}

\institute{School of Computing Science, Newcastle University, UK
           \and PUC-Rio, Rio de Janeiro, Brazil }

%
\title{Tracking Dengue Epidemics using Twitter Content Classification and Topic Modelling
}

\maketitle
\begin{abstract}
Detecting and preventing outbreaks of mosquito-borne diseases such as Dengue and Zika in Brasil and other tropical regions has long been a priority for governments in affected areas.
Streaming social media content, such as Twitter, is increasingly being used for health vigilance applications such as flu detection. However, previous work has not addressed the complexity of drastic seasonal changes on Twitter content across multiple epidemic outbreaks. In order to address this gap, this paper contrasts two complementary approaches to detecting Twitter content that is relevant for Dengue outbreak detection, namely supervised classification and unsupervised clustering using topic modelling. Each approach has benefits and shortcomings. Our classifier achieves a prediction accuracy of about 80\% based on a small training set of about 1,000 instances, but the need for manual annotation makes it hard to track seasonal changes in the nature of the epidemics, such as the emergence of new types of virus in certain geographical locations.
In contrast, LDA-based topic modelling scales well, generating cohesive and well-separated clusters from larger samples. While clusters can be easily re-generated following changes in epidemics, however, this approach makes it hard to clearly segregate relevant tweets into well-defined clusters.

\end{abstract}

\section{Introduction}

Mosquito-borne disease epidemics are increasingly becoming more frequent and diverse around
the globe and it is likely that this is only the early stage of epidemic waves that will continue for several 
decades. Rapidly spreading diseases to combat nowadays are those transmitted by the \textit{Aedes} mosquitoes \cite{Denguecenter2015}, which carry not only \textit{Dengue} virus, but also \textit{Chikungunya} and \textit{Zika} viruses \cite{Denguecenter2015}, which are responsible for thousands of deaths every year. Therefore, improved surveillance through rapid response measures against Aedes-borne diseases is a long-standing tenet 
to various health systems around the world. They are urgently required to mitigating the already 
heavy burden on those health systems and limiting further spread of mosquito-borne diseases within geographical locations, such as in Brazil. Control of Aedes-borne disease requires the vector control by identifying and reducing breeding sites.

Our approach to addressing this problem involves the automatic detection of relevant content in Twitter, in order to determine its relevance as actionable information. 
Paraphrasing \cite{Sakaki2010}, we note that social media users are increasingly viewed as \textit{informative social sensors}, who spontaneously communicate valuable information, which in this case may help in detecting the location and extent of mosquito foci.
However, as the signal produced by these sensors is very noisy, our realistic goal is to automatically categorise Twitter messages into a few classes, segregating recognisable highly informative from less informative and noisy content.

Previous work, e.g. \cite{Gomide2011,Lampos2010,Achrekar2011}, has identified the potential of social media channels, such as Twitter, on offering continuous source of epidemic information, arming public health systems with the ability to perform real-time surveillance. 
However, previously proposed approaches are often limited or insufficient for rapid combat of epidemic waves for several reasons. Firstly, previous work has mainly explored the use 
of social media channels to predict Dengue cases and outbreak patterns by exploring disease-related 
posts from previous outbreaks. However, the combination of socio-economic, environmental and ecological 
factors dramatically changes the characteristics governing each epidemic wave. As a consequence, exploring disease-related posts from previous outbreaks tends to be ineffective to identify breeding
sites in the outset of each outbreak. Secondly, previous work is not aimed at identifying map breeding sites of the mosquito within a region.

\begin{figure*}[htb]
\centering
\includegraphics[width=\linewidth]{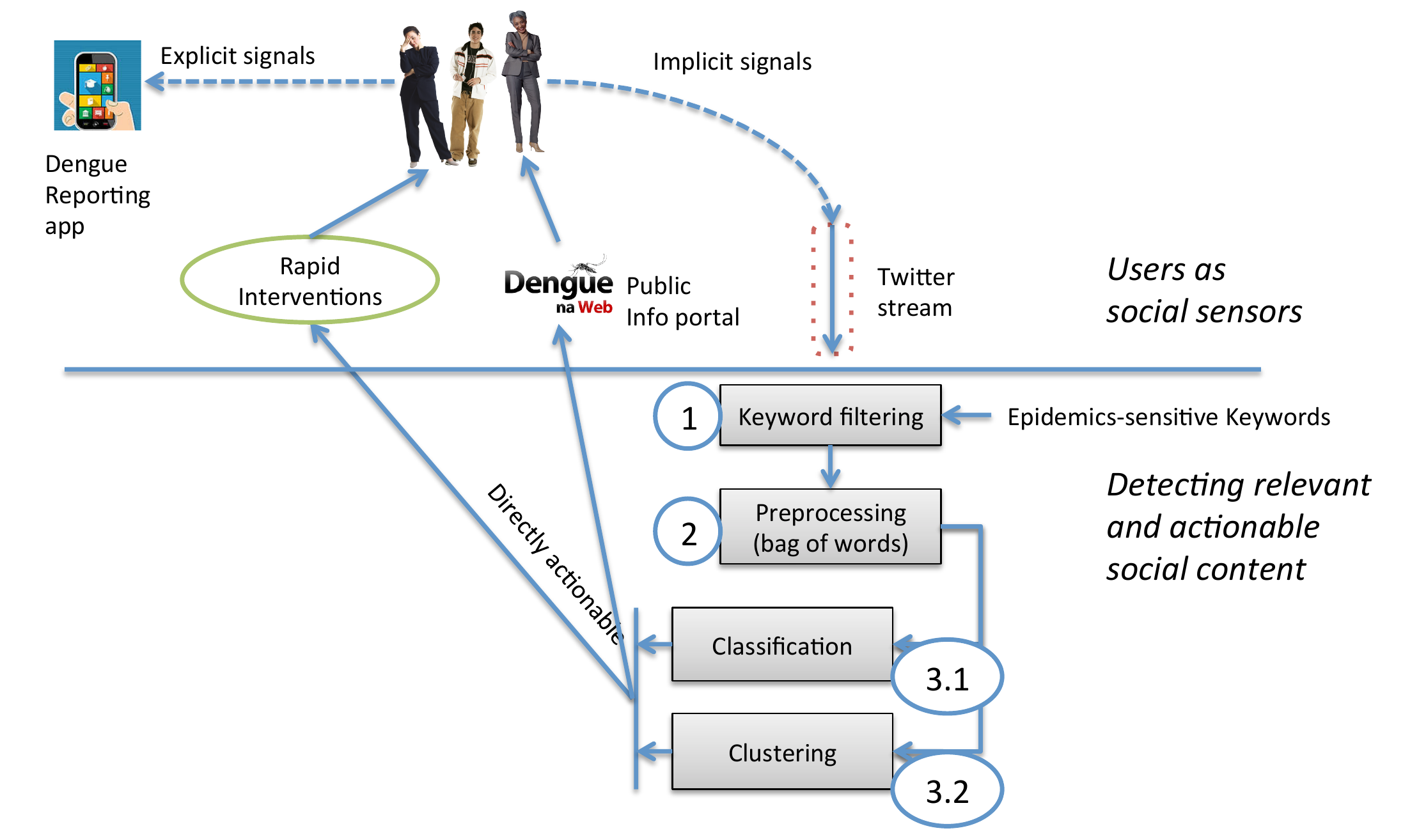}
\caption{Role of automated Twitter relevance detection for health vigilance against Dengue }
\label{fig:Denguestrategy}
\end{figure*}

The role of Twitter content relevance detection is depicted in Fig. \ref{fig:Denguestrategy}.
Social sensors, the people in the upper half of the figure, contribute information either implicitly, i.e., by spontaneously carrying out public conversations on social media channels, or explicitly, i.e., by interacting with dedicate public Web portals and mobile apps. 
As an example, our group in Brazil 
has been developing both such a Dengue mapping portal, and a mobile app that members of the public may use to report cases of Dengue in their local areas \cite{VazaDengue2015}.

As shown in the figure, we monitor the Twitter feed, pre-select tweets according to a broad description of the Dengue topics using keywords, then classify the selected tweets, aiming to segregate relevant signal from the noise. 
We distinguish between relevant signal that is \textit{directly} and \textit{indirectly} actionable. Directly actionable tweets, which we classify as \textit{mosquito focus}, are those that contain sufficient information regarding a breeding site (including geo-location), to inform immediate  interventions by the health authorities.
For instance:
\begin{mdframed}[leftmargin=\parindent,rightmargin=\parindent,skipabove=\topsep,skipbelow=\topsep]  
\scriptsize @Ligue1746 Aten\c{c}\~{a}o! Foco no mosquito da dengue. Av Sta Cruz, Bangu. Em frente ao hospital S\~{a}o Louren\c{c}o! 
(\textit{@Ligue1746 Attention! Mosquito focus found in Santa Cruz avenue, Bangu. In front of the São Lourenço hospital!})
\end{mdframed}
These posts are relatively scarce within the overall stream, however, accounting for about 16\% of the ground truth class assignments.
Indirectly actionable tweets carry more generic information about members of the public complaining about being affected by Dengue (the \textit{Sickness} class), or \textit{News} about the current Dengue epidemics.
For example:
\begin{mdframed}[leftmargin=\parindent,rightmargin=\parindent,skipabove=\topsep,skipbelow=\topsep]  
\scriptsize Eu To com dengue 
(\textit{I have dengue fever})\\
\scriptsize ES  tem mais de 21 mil casos de dengue em 2015
(\textit{ES has more than 21 thousands cases of dengue in 2015})
\end{mdframed}

The rest of the tweets are all considered noise. In particular, these include messages where people joke about Dengue in a sarcastic tone, which is commonly used in online conversation in Brazil, for example:
\begin{mdframed}[leftmargin=\parindent,rightmargin=\parindent,skipabove=\topsep,skipbelow=\topsep]  
\scriptsize Meu WhatsApp ta t\~{a}o parado que vai criar mosquito da dengue 
(\textit{My WhatsAp is so still that it'll create dengue mosquito})
\end{mdframed}

In this paper we report our experiments on automatically classifying directly and indirectly actionable tweets. In the Figure, the classifier plays the role of a filter to limit the amount of noise on the pages displayed on our Web portal.
%

One problem faced in our classification scenario is that \textit{epidemic waves} differ from season to season.
For instance, new symptoms caused by the \textit{Zika} virus have been observed in the epidemic wave, which started in October 2015. 
Such types of epidemic changes drastically change the nature of Twitter content, requiring different keyword settings and filtering from the Twitter feed, in order to accurately track an epidemic. Examples of keywords for tracking different virus and new emerging symptoms include \textit{Dengue}, \textit{Chikunguya}, and, more recently, \textit{Zika}. Simply taking the union of all three would just add to the noise. What is required instead is the ability to rapidly reconfigure the classifier following a drift in topic.
This flexibility requirement naturally suggests an unsupervised approach to learning the classifier. 
At the same time, a supervised classifer that is trained using manually labelled content is likely to be more accurate.

\subsection{Contributions}
In this work we explore the trade-offs between accuracy and flexibility, by comparing and contrasting a supervised classifier learning approach (3.1 in Fig. \ref{fig:Denguestrategy}) with an unsupervised content clustering, using Topic Models (3.2) and specifically on LDA \cite{Blei:2003:LDA:944919.944937}, a popular algorithm that has been previously shown to apply well to clustering Twitter data \cite{ramage2010characterizing,Ritter2012}.
We expect supervised classification to provide good accuracy, as well as give an obvious way to select actionable content from the most informative classes (\textit{mosquito focus}, \textit{sickness}, and \textit{News} in this order). 
On the other hand, this model suffers from known limitations in the size of the training set, which may lead to disappointing performance on content in the wild, and it is expensive to re-train following changes in the filtering keywords.

In contrast, topic modelling is a form of semantic clustering where a clustering scheme can be easily periodically re-generated from large samples. 
While the clear characterisation of clusters using ranked lists of terms from the content's vocabulary (topics) makes this a popular approach, a topic may include heterogeneous content that cuts across expert-defined classes, such as those above, making it harder to associate them with a clear focus. 
This problem is particularly acute in our setting, where we already have a topic defined (through keywords), and we are essentially asking LDA to further refine it in terms of well-separated sub-topics.

We assessed the potential of our approach on large cities of Brazil, such as Rio de Janeiro, by analyzing two cycles of Aedes-related epidemic waves. Our specific contributions in this paper are: (i) a pipeline that implements both methods, including a dedicated pre-processing phase that accounts for idiosincratic use of the Brazilian Portuguese language in tweets, and (ii) an experimental evaluation of their effectiveness.
The supervised classifier is currently in operation as part of the experimental Dengue Web portal developed at PUC-Rio \cite{VazaDengue2015}.

\subsection{Related work}

This paper makes original contributions to an already existing landscape of research on monitoring social media for health vigilance purposes.
Similarly to our work, Twitter data is used by \cite{Gomide2011} to track the Twitter stream and filter relevant signals from it.
Because they only use supervised classification for content filtering, their approach is limited by the amount of labels made available by expert annotators. Moreover, this limitation does enable to easily reveal new information in the outset of each epidemic wave.
In our work, we use not only supervised classification, but also unsupervised clustering as means of identifying relevant social signals.
Finally, we contrast the results from both methods in order to: (i) reflect on the different use cases the methods require (i.e. in terms of annotation effort), and (ii) observe how unsupervised classification helps to better achieve the purpose of revealing new information in each epidemic wave.

\cite{Achrekar2011} and \cite{Lampos2010} show that the frequency of tweets containing simple search keywords can be a good indicator of a trend for a flu epidemic.
The authors show that there is a strong correlation between the number of medically registered visits to a GP concerning flu and the number of tweets mentioning flu.
This approach to tracking epidemics is complementary to ours because, while the previously mentioned authors measure tweet activity on an entire corpus of tweets, we use machine learning to further discover sub-signals in the corpus in specific epidemic waves.
Our approach enables one to further measure and study tweet activity within relevant sub-signals.

Similar methods of monitoring Twitter data have been applied for general event detection, as done, for example, by \cite{Cheng2014} or \cite{Becker2011}.
However, obtaining ground truth is recognised to be a serious bottleneck in a supervised learning pipeline and efforts to reduce the annotation effort have been attempted.
For instance, \cite{Go2009a} automatically identify ground truth from emoticons for sentiment classification. However, as previously mentioned, even if ground truths are somehow identified, the use of supervised learning may not suffice to cope with tracking the changing characteristics of different epidemic waves.

\vspace{-5pt}
\section{Twitter content acquisition and processing}  \label{sec:content}
\vspace{-5pt}
Our experimental dataset consists of three sets of Twitter content, harvested over two periods of time, during the first and second semester of 2015. These periods corresponded to two cycles of epidemic waves. The first two sets, of about 1,000  and 1,600 instances, respectively, were manually annotated by our group at PUC-Rio, which also included the participation of a medical doctor and an epidemiologist. They were used in supervised classification as our training and test set (using standard k-fold validation), and for further testing (no training), respectively.
A larger third set of about 100,000 tweets was used for topic modelling.

A technique similar to that described in \cite{Nagarajan2009} was used to determine a set of filtering keywords for harvesting the tweets.
Namely, we started with the single \#dengue hashtag ``seed'' for an initial collection. 
Upon manual inspection of about 250 initial tweets, our local experts then extended the set to include the most relevant hashtags. These hashtags were those that all local experts agreed to be relevant after discussion amongst them. The final search set contains the following elements (including their common minor variations): 
\{   \#Dengue, \#suspeita, \#Aedes, \#Epidemia, \#aegypti, \#foco, \#governo, \#cuidado, \#febreChikungunya, \#morte, \#parado, \#todoscontradengue, \#aedesaegypti\}.\footnote{Only tweets in the Portuguese language were considered in this study.}

Content pre-processing includes a series of normalisation steps, followed by POS tagging and lemmatisation.\footnote{We used the tagger from Apache OpenNLP 1.5 series (\url{http://opennlp.sourceforge.net/models-1.5/}), and the LemPORT Lemmatizer customised for  Portuguese language vocabulary.}
We normalised the content by removing 38 kinds of ``twitter lingo'' abbreviations, some of which are regional to Brazil (``abs'' for ``abra\c{c}o'', ``blz'' for ``beleza'', etc.), as well as all emoticons and non-verbal forms of expressions.
While those are crucial to understanding the \textit{sentiment} expressed in a tweet, we found that they are not good class predictors, including the \textit{Jokes} class. 
We also replaced links, images, numbers, and idiomatic expressions using conventional terms (\textit{url}, \textit{image}, \textit{funny},...).

%

\vspace{-5pt}
\section{Supervised classification}  \label{sec:supervised}
\vspace{-5pt}

Our classification goal has been to achieve a finer granularity of tweet relevance than just a binary classification into actionable and noise.
The following set of four classes, of decreasing relevance, gave us at the same time a good accuracy and granularity:

\begin{description}
\item[\textit{Mosquito-focus}:] 
this is the most \textit{directly actionable} class, including tweets that report sites that are or may be foci for Dengue mosquito, or sites that provide conducive environments to mosquito breeding. This class accounts for about 16\% of tweets in our test set.

\item[\textit{Sickness}:] This is the second most informative class. These tweets represent cases of: (i) users suspecting or confirming they are sick or they are aware of somebody else who is sick, and (ii) users discussing disease symptoms. Note that previous work (Sect. 1.7) on tracking Aedes-related epidemic waves make no distinction between this and the previous class. 

  \item[\textit{News}:]  This class represents general news about Dengue, ie tweets that spread awareness, report on available preventive measures, inform about health campaigns, and report the number of Dengue cases in certain locations. 
  These are stilll \textit{indirectly actionable} and useful eg. to show emerging outbreak patterns in specific areas.

\item[\textit{Joke}:] Finally, about $20$\% of the tweets in our sample contain a combination of jokes or sarcastic comments about Dengue epidemic. While we regard these as noise, their detection requires an understanding of sarcastic tone in short text, which is challenging as it uses the same general terms as those found in more informative content. 

\end{description}

The training set of about 1,000 messages was annotated by three local experts independently, by taking the majority class for each instance, requiring about 100 hours over three refinement steps to resolve inconsistencies and ambiguities.
The classes are fairly balanced:
\textit{News}: 333  (31\%), 
\textit{Joke}: 148 (14\%), \textit{Mosquito focus}: 257 (24\%), and \textit{Sickness}: 338 (31\%). 
Classification performance, measured using standard cross-validation, was similar across different classifier models, namely SVM, Naive Bayes, and MaxEntropy.
We chose Naive Bayes as having probabilities associated to each class assignment helped identify the weak assignments, and thus the potential ambiguities in the manual annotations.

The classifier reported an overall 84.4\% accuracy and .83 F-measure.
In order to further validate these results, we then sampled an additional set of 1,600 tweets, none of them used for training, and performed both automated classification and manual annotation on this set.
On this new set, the distribution of instances in each class, taken from the ground truth annotations, is not substantially different from that in the training set, except for the more abundant \textit{Mosquito focus} class: \textit{News}: 404 (25\%), 
\textit{Joke}: 289 (18\%), \textit{Mosquito focus}: 253 (16\%), and \textit{Sickness}: 649 (41\%). Performance results for this classifier are reported in Table  \ref{tab:supervised-results}.
%

\begin{table}
\centering
\begin{footnotesize}
\begin{tabular}{|c|c|c|c|c|}
\hline \textbf{Class}  & \textbf{Precision} & \textbf{Recall} & \textbf{F} & \textbf{Accuracy} \\
\hline \textbf{News}  & .79 & .74 & .76 &  .74 \\
\textbf{Joke}             & .63 & .85 & .72 & .85 \\ 
\textbf{Mosquito focus}  & .79 & .85 & .83 & .86 \\ 
\textbf{Sickness}         & .91 & .78 & .84 & .78 \\ 
\hline 
\end{tabular}
\end{footnotesize}
\caption{Classifier performance on independent test set}
\label{tab:supervised-results}
\end{table}
\vspace{-15pt}

\section{Unsupervised content clustering using LDA}  \label{sec:unsupervised}

As discussed earlier, supervised classification does not fully meet our requirements, as manual annotation limits the size of training set and makes it difficult to update the model when the characteristics of the epidemics changes.  Also, finding a crisp, unambiguous classification has been problematic.

LDA-based clustering \cite{Blei:2003:LDA:944919.944937} has been used before for Twitter content analysis and topic discovery, for example by \cite{Morstatter2013,Rosa2011,Weng2010}.
What we investigate is an application of LDA that shows the potential for scalability and flexibility, i.e., by periodically rebuilding the clusters to track drift in Twitter search keywords.

For this experiment, our sample dataset consists of $107,376$ tweets, harvested in summer 2015 using standard keyword filtering from the Twitter feed, and containing a total of $17,191$ unique words.
Raw tweets were pre-processed just like for classification (phase 2 in Fig. \ref{fig:Denguestrategy}), 
producing a bag-of-words representation of each tweet. 
Additionally, as a further curation step we removed the $20$ most frequent words in the dataset, as well as all words that do not recur in at least two tweets.
This last step is needed to prevent very frequent terms from appearing in all topics, which reduces the effect of our cluster quality metrices and cluster intelligibility.

\subsection{Evaluation of clustering quality}


We explored a space of clustering schemes ranging from 2 to 8 clusters.\footnote{All experiments carried out using the Apache Spark LDA package \small \url{https://spark.apache.org}}
In the absence of an accepted gold standard, a number of evaluation methods have been proposed in the literature.
For instance, \cite{Morstatter2013} proposes to measure cluster quality by quantifying the differences caused in topic mining using two different stream sampling mechanisms.
The method is based on the differences between the distribution of words across topics and between the two sampling mechanisms. 
However, it cannot be used in our setting, because our corpus of tweets is fixed, rather than a sample.
Also, while any two individual words may have different frequency distributions, the approach does not necessarily take into account the importance, measured by relative frequency, of the words within the entire corpus.
In an alternative approach, \cite{Rosa2011} use ground truth in the form of pre-established hashtags.
This is not applicable in our scenario, either, because by the way our topic filtering is done, most of the tweets in our corpus will already include a high number of hashtags, including for instance the \#dengue hashtag.

Instead, we propose to use \textit{intra-} and \textit{inter-} cluster similarity as our main evaluation criteria.
This is inspired by \textit{silhouettes} \cite{Rousseeuw1986}, and based on the contrast between \textit{tightness} (how similar data are to each other in a cluster) and \textit{separation} (how dissimilar data are across clusters).
Specifically, we define the similarity between two clusters $C_a, C_b$ in terms of the cosine TF-IDF similarity of each pair of tweets they contain, i.e., $t_i \in C_a$ and $t_j \in C_b$, as follows:
\begin{equation}
	\mathit{sim}(C_a, C_b) = \frac{1}{|C_a| \; |C_b|} \sum_{t_i \in C_a, t_j \in C_b}\frac{\mathbf{v}(t_i) \cdot \mathbf{v}(t_j)}{||\mathbf{v}(t_i)|| \; ||\mathbf{v}(t_j)||}
\label{eq:similarity}
\end{equation}
where $\mathbf{v}(t_i)$ is the TF-IDF vector representation of a tweet. That is, 
the $k$th element of the vector, $t_i[k]$, is the TF-IDF score of the $k$th term. As a reminder, the TF-IDF score of a term quantifies the relative importance of a term within a corpus of documents \cite{Aggarwal2012}. Eq. (\ref{eq:similarity}) defines the \textit{inter-cluster similarity} between two clusters $C_a \neq C_b$, while the \textit{intra-cluster similarity} of a cluster $C$ is obtained by setting $C_a = C_b = C$.

\begin{figure}
	\centering
	\includegraphics[width=.8\textwidth]{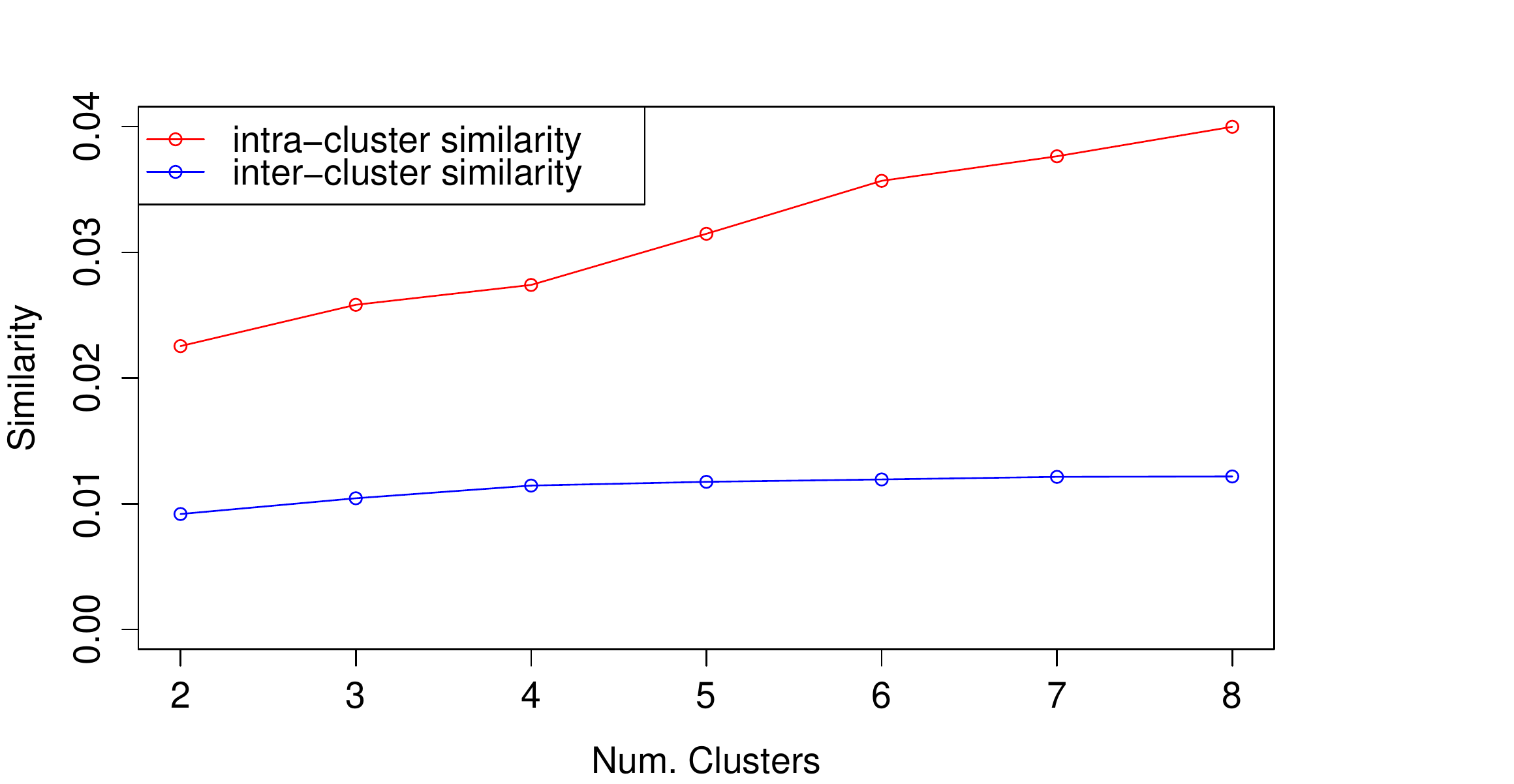}
	\caption{Intra- and Inter-cluster similarities}
	\label{fig:intra.inter}
\end{figure}

Fig.\ \ref{fig:intra.inter} reports the inter- and intra-cluster similarity scores for each choice of clustering scheme.
The absolute similarity numbers are small,  due to the sparse nature of tweets and the overall little linguistic overlap within clusters.
However, we can see that the intra-cluster similarity is more than twice the inter-cluster similarity, indicating good separation amongst the clusters across all configurations.
This seems to confirm that the LDA approach is sufficiently sensitive to discover sub-topics of interest within an already focused general topic, defined by a set of keywords.

\begin{figure*}[ht]
\centering
\begin{minipage}[b]{0.5\linewidth}
	\includegraphics[width=\linewidth]{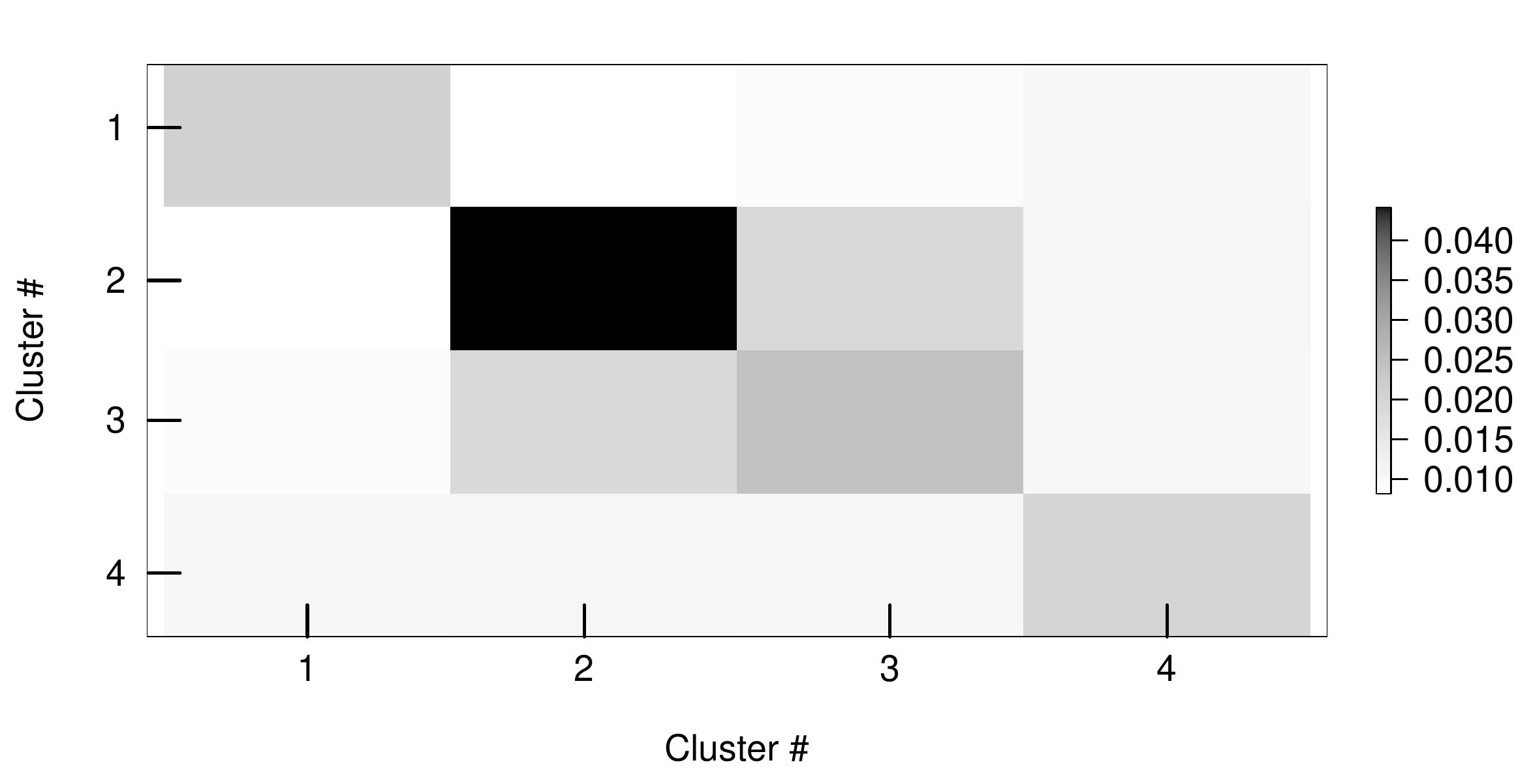}
\end{minipage}
\quad
\begin{minipage}[b]{0.45\linewidth}
	\includegraphics[width=\linewidth]{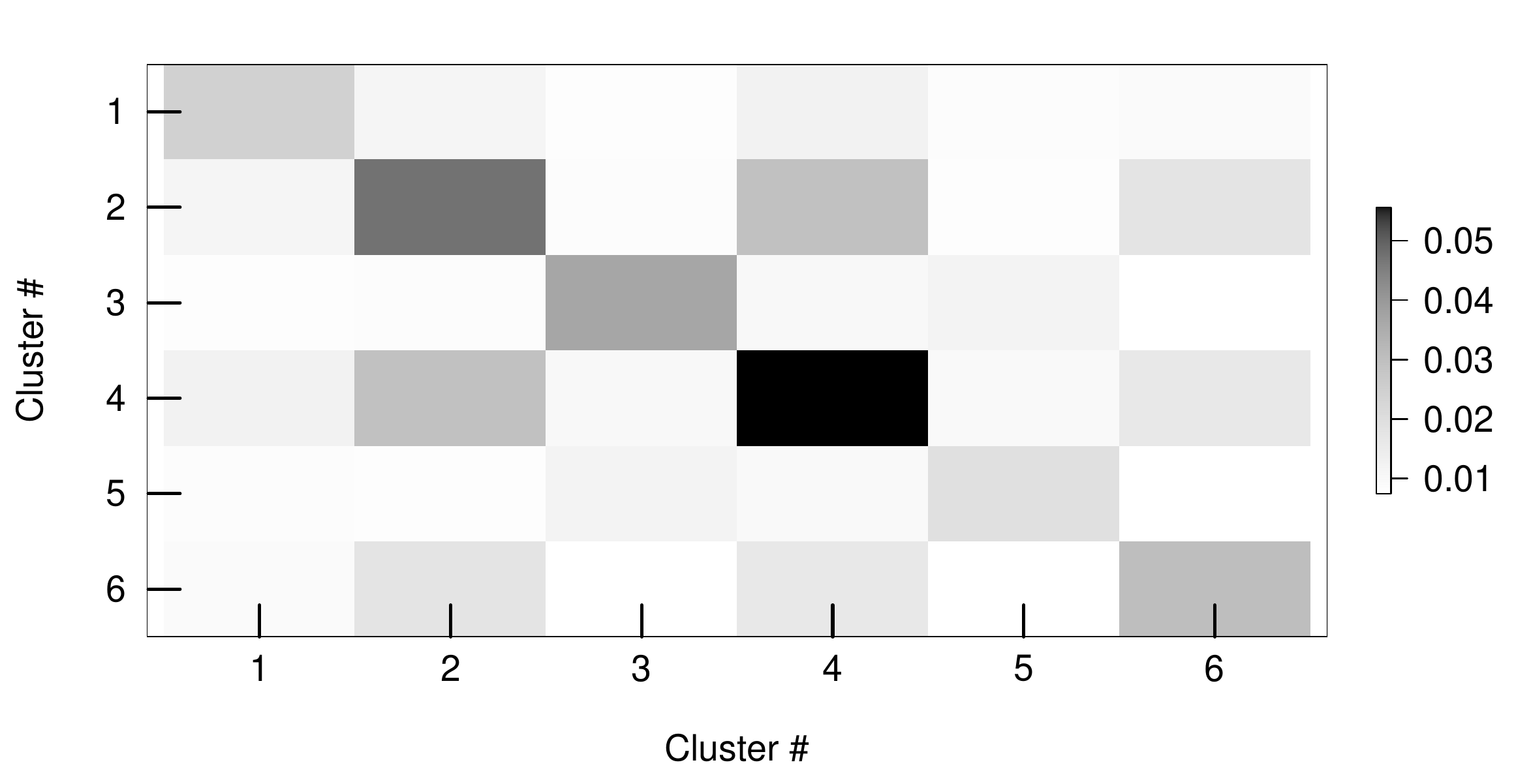}
\end{minipage}
	\caption{Inter- and intra-similarity for 4 and 6 clusters topic models}
	\label{fig:clusters}
	\end{figure*}

The plots in Fig.\ \ref{fig:clusters} provide more detailed indication of the contrast between intra- and inter-cluster similarity at the level of detail of individual clusters.
For example, in the 4-clusters case, the average of the diagonal values of the raster plot is the intra-cluster similarity reported in Fig.\ \ref{fig:intra.inter}, whereas the average of the off-diagonal values represent the inter-cluster similarity. In these plots, darker boxes indicates higher (average) similarity. 
Thus, a plot where the diagonals are darker than the off-diagonal elements is an indication of a high quality clustering scheme.

%
%
%
%
%
%
%

Although the similarity metrics are objective and seems to confirm the good quality of the clustering, the plots in Fig. \ref{fig:intra.inter} and Fig. \ref{fig:clusters} do not provide much insight into the optimal number of clusters, or indeed their semantic interpretation.
We therefore relied on our domain experts for the empirical selection of the clustering scheme (2,4,6,8 clusters) that would most closely lend itself to an intuitive semantic interpretation of the topics.
Their assessment is reported below.
In Sec. \ref{sec:comparison} we present a comparison of topics content using our four classes model as a frame of reference.

\subsection{Empirical topics interpretation}

Expert inspection, carried out by native Brazilian Portuguese speakers, considered both the list of words within each topic, and a sample of the tweets for that topic.
In this case, the most intelligible clustering scheme had $4$ topics.
The following is a list of most relevant topics for this scheme:
\begin{mdframed}[leftmargin=\parindent,rightmargin=\parindent,skipabove=\topsep,skipbelow=\topsep]  
\scriptsize 	\textbf{Topic 1:} parado, 	\'{a}gua, fazer, vacina, 	at\'{e}, meu, 	t\~{a}o \\
    \textbf{Topic 2:} combate, morte, s\'{a}ude , confirma, a\c{c}\~{a}o , homem, chegar, queda, confirmado, agente \\
	\textbf{Topic 3:} contra, suspeito, s\'{a}ude , doen\c{c}a, bairro, morrer, combater, cidade, dizer, mutiro \\
	\textbf{Topic 4:} mosquito, epidemia, pegar, foco, casa, hoje, mesmo, estado, igual
\end{mdframed}
%
	 
%
%
%
The importance of the words is given by LDA as a measure of how well they are represented in the topics.\footnote{Some of the words are just noise. This is due to occasional imperfect lemmatisation during the preprocessing stage.}
Unsurprisingly, topic inspection suggests an interpretation that only partially overlaps with the a priori classification we have seen in the supervised case.
Specifically, \textbf{Topic 1} is closely related to \textit{Jokes}.
Most of the tweets for this topic either make an analogy between Dengue and the users lives, or they use the words related to Dengue as a pun.
A typical pattern is the following:
\begin{mdframed}[leftmargin=\parindent,rightmargin=\parindent,skipabove=\topsep,skipbelow=\topsep]  
\scriptsize meu [algo como: wpp - WhatsApp, timeline, Facebook, twitter etc] est\'{a} mais parado do que agua com dengue.\\
\textit{My [something like: wpp - WhatsApp, timeline, Facebook, twitter etc] is more still than standing water with dengue mosquito.}
\end{mdframed}
Specific examples include:
\begin{mdframed}[leftmargin=\parindent,rightmargin=\parindent,skipabove=\topsep,skipbelow=\topsep]  
\scriptsize Aitizapi ta com dengue de t\~{a}o parado  (\textit{Aitizapi is so still that it has been infected by dengue})\\
\scriptsize Concession\'{a}ria t\'{a} dando dengue de t\~{a}o parada que t\'{a} (\textit{Car dealership is so still that it has dengue})
\end{mdframed}

In the first example, the user was playing with the words when referring to the standing status and inactivity in his Whatsapp account. Breeding sites of the Aedes mosquito are mostly found in containers with standing water.
In the second, the user is joking about significant decreases in car purchases due to the emerging economic crisis in Brazil. Many of the jokes in the last epidemic wave have been related to Zika, which in Braziliasn Portuguese, has been used as a new slang word for failure or any kind of personal problem.    

\textbf{Topic 2} is interpreted as \textit{news} about increase or decrease of Aedes-borne disease cases as well as specific cases of people who died because of the Aedes-borne diseases, i.e. Dengue, Chikungunya and Zika. It also contains news about the combat of the mosquito in certain locations as well. Examples:
\begin{mdframed}[leftmargin=\parindent,rightmargin=\parindent,skipabove=\topsep,skipbelow=\topsep]  
\scriptsize Rio Preto registra mais de 11 mil casos de dengue e 10 mortes no ano  \#SP\\
\textit{Rio Preto reports more than 11 thousand cases of dengue in the year \#SP} \\
\\
\scriptsize 543 casos est\~{a}o em an\'{a}lise - Londrina confirma mais de 2,5 mil casos de dengue em 2015 - [URL removed]\\
\textit{543 cases of dengue are under analysis - Londrina confirms more than 2.5 cases of dengue in 2015 - [URL removed]}
\end{mdframed}

\textbf{Topic 3} appears to contain mostly \textit{news about campaigns} or actions to combat or to prevent Aedes-borne diseases, for instance:
\begin{mdframed}[leftmargin=\parindent,rightmargin=\parindent,skipabove=\topsep,skipbelow=\topsep]  
\scriptsize Curcuma contra dengue [URL removed] (\textit{Curcuma against dengue})\\
\\
\scriptsize Prefeitura de Carapicuíba realiza nova campanha contra dengue e CHIKUNGUNYA[URL removed] \\
\textit{Carapicuíba City Hall launches new campaign against dengue and CHIKUNGUNYA[URL removed]}
\end{mdframed}

The difference between the news in topics 2 and 3 concerns the type of news, which for topic 2 is mostly about the increase or decrease of Aedes-borne diseases, whereas in topic 3 is about campaigns or actions to combat the propagation of the Aedes mosquito.

Finally, \textbf{Topic 4} contains mostly \textit{sickness} tweets, with some instances of \textit{jokes}:

\begin{mdframed}[leftmargin=\parindent,rightmargin=\parindent,skipabove=\topsep,skipbelow=\topsep]  
\small Ser\'{a} que eu to com dengue ? (\textit{I wonder: do I have dengue?})
\end{mdframed}


\subsection{Classes vs clusters}  \label{sec:comparison}

The point to note in the assessment above is that the most relevant tweets, those corresponding to the \textit{Mosquito Focus} class, are not easily spotted, in particular they do no seem  characterise  any of the topics.
Intuitively, this can be explained in terms of the relative scarcity of these tweets within the stream, combined with the balancing across topics that occurs within LDA. 

In order to quantify this intuition, we have analysed the topics content using our pre-defined four classes as a frame of reference.
In this analysis, we have used our trained classifier to predict the class labels of all the tweets in the corpus that we used to generate the topics (about 100,000). 
We then counted the proportion of class labels in each topic, as well as, for each class, the scattering of the class labels across the topics.
The results are presented in Table \ref{tab:classesByCluster} and Table \ref{tab:classesAcrossClusters}, respectively, where the dominant entries for each column (resp row) are emphasised.

\begin{table*}[ht]
\centering
\begin{minipage}[b]{0.45\linewidth}
\begin{scriptsize}
\begin{tabular}{|c|c|c|c|c|}
\hline  &  Topic 1 &  Topic 2 & Topic 3  & Topic 4  \\ 
\hline  News &  13.9 &  \textbf{72.6} &  27.2 &  \textbf{39.4} \\ 
\hline  Joke &  \textbf{39.5} &  0.1 &  2.8 &  4.1 \\ 
\hline  Mosquito Focus &  30 &  4.0 &  12.3 &  12.5 \\ 
\hline  Sickness &  16.6  &  23.3 &  \textbf{57.7} &  \textbf{44.0} \\ 
\hhline{|=|=|=|=|=|}  Total  &  100 &  100 &  100 & 100  \\ 
\hline 
\end{tabular} 
\end{scriptsize}
\caption{Distribution (\%) of predicted class labels within each cluster}
\label{tab:classesByCluster}
\end{minipage}
\quad
\begin{minipage}[b]{0.45\linewidth}
\begin{scriptsize}
\begin{tabular}{|c|c|c|c|c||c|}
\hline  &  Topic 1 &  Topic 2 & Topic 3  & Topic 4 & Total  \\ 
\hline  News &  29.1 &  28.5 &  8.9 &  33.5  & 100 \\ 
\hline  Joke &  \textbf{95.0} &  0.03 &  1.05 &  4.0   & 100 \\ 
\hline  Mosquito Focus &  \textbf{79.5} &  2.0 &  5.1 &  13.4   & 100 \\ 
\hline  Sickness &  \textbf{34.8}  &  9.1 &  18.8 &  \textbf{37.3}   & 100 \\ 
\hline 
\end{tabular}
\end{scriptsize}
\caption{Scattering (\%) of predicted class labels across clusters}
\label{tab:classesAcrossClusters} 
\end{minipage}
\end{table*}
\vspace{-15pt}

It is worth remembering that these results are based on predicted class labels and are therefore inherently subject to the classifier's inaccuracy.
Furthermore, the predicted class labels were \textit{not} available to experts when they inspected topic content, thus they effectively performed a new manual classification on a content sample for each topic.
Despite the inaccuracies introduced by these elements, Table \ref{tab:classesByCluster} seems to corroborate the experts' assessment regarding topics 1 and 2, but less so for topics 3 and 4.
This may be due to the sampling operated by the experts, which selected content towards the top of the topic (LDA ranks content by relevance within a topic) and may have come across joke entries which are otherwise scarce in topic 4.
Although the heavy concentration on joke tweets in topic 1 from Table \ref{tab:classesByCluster} seems promising (i.e., the other topics are relatively noise-free), Table \ref{tab:classesAcrossClusters} shows a problem, namely that topic 1 is also where the vast majority of \textit{Mosquito Focus} tweets are found. 
Thus, although topic 1 segregates the most informative tweets well, it is also very noisy, as these tweets are relatively scarce within the entire corpus.

The analysis just described suggests that topic modelling offers less control over the content of topics, compared to a traditional classifier, especially on a naturally noisy media channel. 
Although relevant content can be ascribed to specific topics, these are polluted by noise.
Despite this, LDA performs relatively well on creating sub-topics from a sample that is already focused on a specific topic, such as conversations on the Aedes-transmitted viruses.
The main appeal of the classifier is that it makes it straightforward to select relevant content, with acceptable experimental accuracy.
In our follow on research we are investigating ways to combine the benefits of the two approaches.
Specifically, we are studying a unified semi-supervised model where topic modelling can be used to improve the accuracy of the classifier, i.e., by automatically expanding the training set, and to alleviate the cost of re-training at the same time.

\vspace{-10pt}
\section{Summary}  \label{sec:conclusions}
\vspace{-5pt}
In this paper we discussed methods for detecting relevant content in a Twitter stream that has been pre-filtered to focus on a specific topic, in this instance online discussions around Dengue and other Aedes-borne diseases in Brazil.
Relevance is defined operationally in terms of four classes within the broad topic of Dengue.
When reliably segregated from noise, relevant content can be used in multiple ways in the context of health vigilance to combat epidemics caused by the Aedes mosquito.
We have compared two approaches for detecting relevance, supervised classification and clustering by topic modelling. 
Our experimental results indicate that the clusters produced using topic modelling tend to be noisy, perhaps because LDA is not very effective on text content that is pre-filtered for a specific set of keywords.
Supervised classification, on the other hand, is costly as manual annotation requires multiple rounds due to ambiguities in the content, but is more appealing as a good proportion of actionable messages are segregated, i.e., in the two most relevant classes.
We are currently exploring ways to combine the two approaches into one semi-supervised model, i.e., by exploiting the topics to enhance the training set and alleviate the cost of re-training.

\section*{Acknowledgments}
\vspace{-10pt}

\begin{footnotesize}
This work has been supported by MRC UK and FAPERJ Brazil within the Newton Fund Project entitled \textit{A Software Infrastructure for Promoting Efficient Entomological Monitoring of Dengue Fever}. The authors would like to thank Oswaldo G. Cruz (Fundação Oswaldo Cruz, Programa de Computacao Cientifica) and Leonardo Frajhof (Unirio, Rio de Janeiro, Brazil) for their contributions to this paper, and Prof. Wagner Meira Jr. and his team for sharing their 2009-2011 Twitter datasets \cite{Gomide2011}.
\end{footnotesize}

\vspace{-10pt}
\input{Dengue-tweets-classification-SoWeMine.bbl}

\end{document}

%% file: preamble.tex
\usepackage{times}
\usepackage{helvet}
\usepackage{courier}
\usepackage{url}
\usepackage{graphicx}
\usepackage{mdframed}
\usepackage{hhline}

\usepackage{xcolor}
\definecolor{shadecolor}{rgb}{0.9,0.9,0.9}
\newcommand{\comment}[1]{}
\definecolor{Orange}{rgb}{1,0.5,0}

%% file: Dengue-tweets-classification-SoWeMine.bbl
\newcommand{\etalchar}[1]{$^{#1}$}